\documentclass[reprint,superscriptaddress,amsmath,amssymb,prl]{revtex4-1}

\usepackage{graphicx}% Include figure files
\usepackage{subfigure}% Include subfigure 
\usepackage{dcolumn}% Align table columns on decimal point
\usepackage{bm}% bold math
\usepackage{xcolor}%
\usepackage{amsmath}% useful mathematical things like underbraces
\usepackage{verbatim}% comments
\usepackage{mathtools}
\usepackage{epstopdf}
\usepackage[normalem]{ulem}

\def\nin{\noindent}
\def\beq{\begin{equation}}
\def\eeq{\end{equation}}

\newcommand{\blue}[1]{\textcolor{blue}{#1}}

\begin{document}

\title{The origin of detailed-balance steady states in \\
non-equilibrium dynamics of granular matter}

   \author{Clara C. Wanjura}\email{clara.wanjura@mpl.mpg.de}
   \affiliation{Cavendish Laboratory, Cambridge University,
                JJ Thomson Avenue, Cambridge CB3 0HE, UK}
   \address{Max Planck Institute for the Science of Light, Staudtstraße 2, 91058 Erlangen, Germany}
   \author{Amelie Mayl\"ander}
   \affiliation{Ulm University, Albert-Einstein-Allee 11, 89081 Ulm, Germany}
   \author{Othmar Marti}
   \affiliation{Ulm University, Albert-Einstein-Allee 11, 89081 Ulm, Germany}
 \author{Raphael Blumenfeld}\email{rbb11@cam.ac.uk}
    \affiliation{Gonville \& Caius College, Cambridge University,
                Trinity St., Cambridge CB2 1TA, UK}
   \affiliation{Imperial College London, London SW7 2AZ, UK}

\date{\today}

\begin{abstract}
Modelling the dynamics of dense granular media is a long standing challenge and essential to many natural phenomena and technological applications.
Here, we trace back puzzling experimental observation of detailed-balanced steady states to self-organisation of the neighbour probability distribution.
The emergence of detailed balance in non-equilibrium granular dynamics could constitute a major step toward better models of granular media, as well as provide more insight into non-equilibrium processes in general.
We show analytically that DBSS emerges when a certain neighbour probability is uniform across the system. This condition leads to a conditional cell order distribution being independent of the condition. We then carry out rotational shear experiments, in which this condition is satisfied, and show that they give rise to robust detailed-balanced steady states. 
We also show that, when the unconditional cell order distribution maximises the entropy, it is determined by a single constant parameter that is characteristic of all cell transitions.
These results illustrate the predictive power of recently proposed evolution equations, which pave the way to simpler models of the dynamics of planar granular systems.

%% emphasise that we also perform experiments

\end{abstract}


\maketitle

Granular matter is ubiquitous in nature and important to most life on Earth, as well as to technological applications. Modelling of granular dynamics is hindered by their out-of-equilibrium nature, which vitiates many established equilibrium-based approaches. In particular, the structures that the dynamics settle into, determine many of their large-scale properties.
Recent experiments~\cite{Sun2021} may have opened the door to use such approaches. Following an earlier conjecture~\cite{Wanjura2020}, they discovered that quasistatic shear of planar granular systems gives rise to steady states that robustly satisfy detailed balance of certain elementary processes. These observations are puzzling in view of the accepted paradigm that detailed balance is only a feature of thermal equilibrium~\cite{Klein1955}. 
Since detailed balance played a key role in physics modelling~\cite{Ma1867,Bo1872,Ein17,Di24,On31}, its emergence in this context could be instrumental for similar progress in granular science and potentially for other non-equilibrium dynamics.

\begin{figure}[htbp]
   \centering
   \includegraphics[width=.45\textwidth]{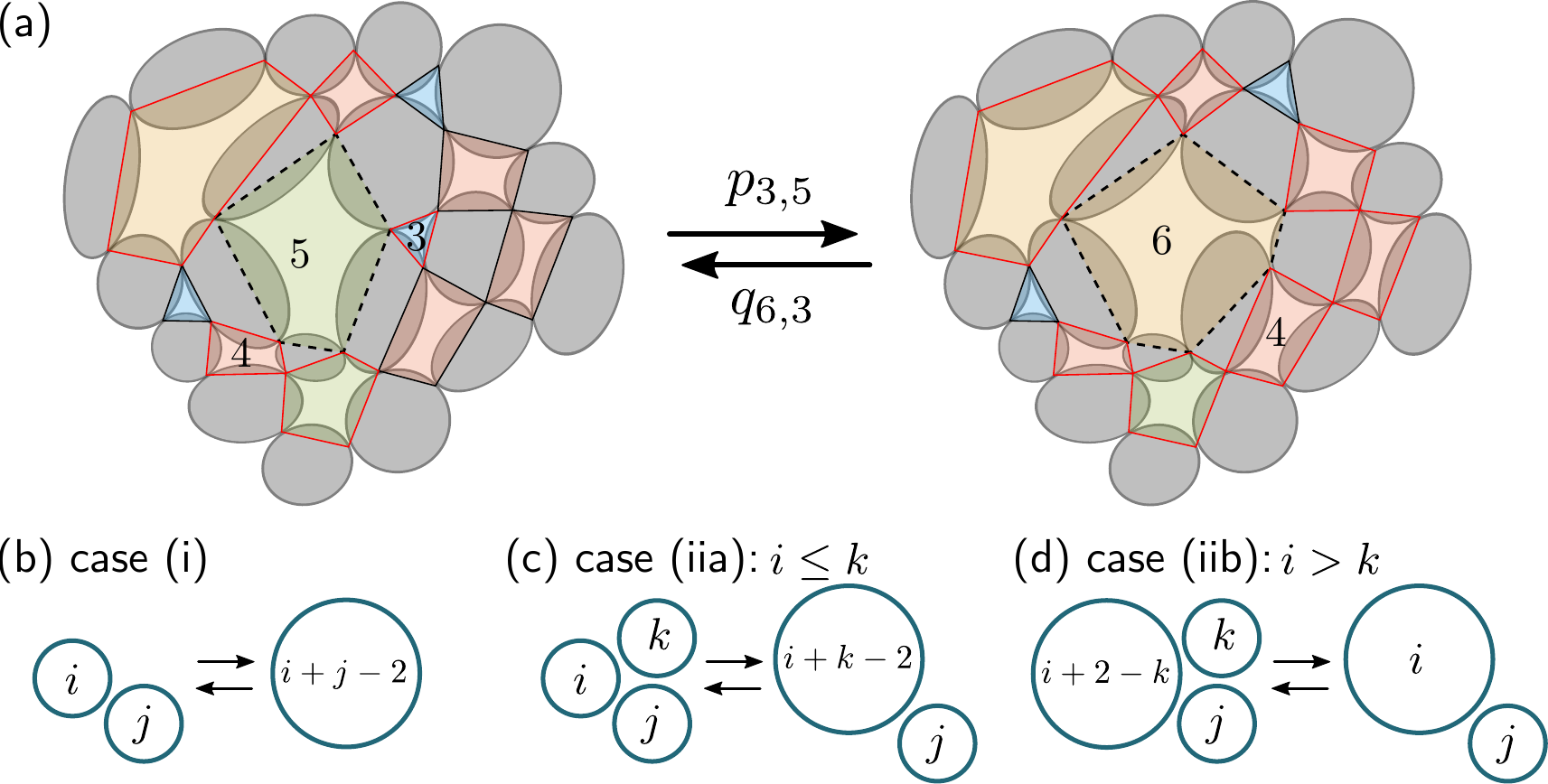}
   \caption{\textbf{Cell order transitions.}
   (a)~The merging of a $3$- and a $5$-cell into a $6$-cell by contact breaking. We show the neighbours of the central $5$-cell (dashed line) on the left and the $6$-cell on the right in red.
   (b)-(d)~Different cases contributing to a change in the number of $i$-$j$ neighbours.
   }
\label{fig:CE}
\end{figure}

In planar systems, the structure can be partitioned into 'cells', which are the smallest loops enclosed by particles in contact. It is useful to define a cell order as the number of particles in contact surrounding it~\cite{Bagi96}. Quasistatic dynamics of such systems progress through particle movements that break and make inter-granular contacts.
Breaking a contact merges the two cells sharing it and making a contact splits a cell of order $k$ (henceforth $k$-cell) into two (Fig.~\ref{fig:CE}(a)).
The equations for the evolution of the cell order distribution (COD), $Q_k(t)$~\cite{MaBl14,MaBl17}, proposed in \cite{Wanjura2020}, predict that, in very dense granular systems, the steady-state merging and splitting events satisfy detailed balance. This prediction, which depended only on the assumption of homogeneity of $Q_k$ is intriguing because it is commonly accepted that: (i)~only systems in thermal equilibrium satisfy detailed balance and (ii)~that balance in steady states of non-equilibrium systems can only be achieved via cyclic kinetic processes, e.g., of the form $A\to B\to C\to A$~\cite{Klein1955}. The prediction in \cite{Wanjura2020} contradicts the first statement trivially because it was valid only for very dense systems that have no cyclic processes anyway~\cite{footnoteCycles}. Moreover, such systems are almost impossible to realise experimentally, making this prediction difficult to test.
Nevertheless, independent experiments~\cite{Sun2021} found that cyclically sheared, much less dense granular systems also exhibit detailed-balanced steady states (DBSS).
Finding such states in any granular system suggests that some aspects of conventional statistical mechanics can be used to model them~\cite{EdOa89a}.
Further support of the usefulness of statistical mechanics to granular systems is a recent finding that entropy plays a key role in the quasistatic structural organisation of granular systems~\cite{Sun2020}. They found that, in systems with sufficiently high inter-particle friction, the steady state COD maximises the entropy, which corresponds to maximising the number of possible cell configurations under given constraints.

Here, we first show analytically that DBSS {\it must emerge} in quasistatically evolving granular systems when the conditional COD around any $k$-cell, $Q_{i\mid k}$, is independent of $k$. 
Secondly, we show that this condition is met in our experiments, reported here, as it was in the experiments in~\cite{Sun2021}. We claim that this is the reason for the observations of DBSS in these experiments.
Thirdly, we show that, if the steady-state COD also maximises the entropy, as found in some experiments~\cite{Sun2021}, then the ratio of cell splitting and merging rates is constant for any such process and it determines the COD.
Finally, we support our results with experiments of rotationally-sheared systems.
Our results show that the number of parameters required to model quasistatic dynamics of granular matter can be reduced significantly, paving the way to simpler models.

\section{The cell order distribution and the steady state}
The COD's evolution equations, in a planar granular system of $N\gg 1$ particles, are~\cite{Wanjura2020}
\begin{align}
   \dot{Q}_k = 
       & \frac{1}{2} \sum_{i=3}^{k-1} \eta_{i,k-i+2} (1+\delta_{i,k-i+2}) \notag \\
       & -\sum_{i=k+1}^{C-2} \eta_{k,i-k+2} (1+\delta_{k,i-k+2})
       + Q_k \sum_{\substack{i,j=3\\i\leq j\\i+j-2\leq C}} \eta_{i,j} \ .
       \label{eq:masterEqs}
\end{align}
Here, $\eta_{i,j}\equiv p_{i,j}Q_iQ_{j\mid i} - q_{i+j-2,i}Q_{i+j-2}$ is a `balance parameter'--- a measure of the mean direction of the process $i+j\leftrightharpoons(i+j-2)$. 
The similarity of this process to a chemical reaction prompts us to call it a `reaction' in the following. $C$ is the highest possible
cell order in the system, $p_{i,j}$ is the rate of $i$- and $j$-cells merging into $(i+j-2)$-cells, $q_{i,i+j-2}$ the rate of splitting of $(i+j-2)$-cells into $i$- and $j$- cells, the $\delta$-functions ensure correct counting, and the rightmost sum accounts for the fluctuating number of cells. In these equations, the COD is assumed to be spatially homogeneous. This assumption will be discussed in detail below. Particles that do not transmit forces in the absence of external force fields (rattlers) are omitted in Eq.~(\ref{eq:masterEqs}), because they have been shown to affect negligibly quasistatic simple-shear dynamics~\cite{Sun2020}. 

At steady state, $\dot{Q}_k=0$ for all $k$. One possible such state is $\eta_{i,j}=0$ for all $i,j$, which is the DBSS, $i+j\leftrightharpoons(i+j-2)$. When $C\geq6$, kinetic cycles are possible, in principle~\cite{footnoteCycles}. In non-equilibrium thermal systems, such cycles must occur, prohibiting detailed balance~\cite{Klein1955}. The question we address next is why DBSS emerged in the experimental a-thermal systems in \cite{Sun2021}. 

The probability that $i$- and $j$-cells are neighbours at time $t$ is the fraction of the number of such pairs, $\mathcal{N}_{i,j}$, out of all possible pairs, $\mathcal{N}=\bar{z}N$/2, where $\bar{z}$ is the mean number of contacts per particle:
\begin{align}
   P_{i,j}(t) = \frac{2\mathcal{N}_{i,j}(t)}{\bar z N} \ .
 \label{eq:neighbourProb}
\end{align}
Here, $N$ is the total number of particles and boundary corrections have been neglected.
As illustrated in Figs.~\ref{fig:CE}(b)-(d), the number of pairs, $\mathcal{N}_{i,j}$, can change in two ways during the reaction, $k+\ell\to (k+\ell-2)$ ($3\leq k,\ell, k+\ell-2\leq C$): (i)~both $i$- and $j$-cells are involved, e.g., $i=k$ and $j=\ell$; (ii)~only $i$ or $j$ is involved.
In (i), each such reaction changes $\mathcal{N}_{i,j}$ by $\pm1$, depending on the reaction direction. For example, in Fig.~\ref{fig:CE}, $i=k=3$, $j=\ell=5$, and $3+5\to6$, depleting $\mathcal{N}_{3,5}$ by one. The net rate of this reaction is $\eta_{i,j}$ and, therefore, it contributes $-N_\mathrm{c}\eta_{i,j}$ to $\dot{\mathcal{N}}_{i,j}$, with $N_\mathrm{c}$ the total number of cells.
Case (ii) comprises two sub-cases: (iia)~$i+k\leftrightharpoons{} (i+k-2)$ ($k\neq j$, $3\leq k\leq C+2-i$), illustrated in Fig.~\ref{fig:CE}(c); (iib)~$k+(i+2-k)\leftrightharpoons{} i$ ($k\neq j$, $3\leq k\leq C$), illustrated in Fig.~\ref{fig:CE}(d). In each sub-case, $\mathcal{N}_{i,j}$ and $N_\mathrm{c}$ change by $\pm1$. The net contribution of these events to $\dot{\mathcal{N}}_{i,j}$ depends on the difference between the reaction rates, $\sum_\ell\eta_{\ell,i+2-\ell} - \sum_k\eta_{i,k}$, and on the number of $i$-$j$ neighbours, $\mathcal{N}_{i,j}=P_{i,j}\bar{z}N/2$.  E.g., in Fig.~\ref{fig:CE}, the reaction $3+5\to6$ contributes $-\bar{z}NP_{4,5}\eta_{3,5}/2$ to $\dot{\mathcal{N}}_{4,5}$ and $+\bar{z}NP_{4,6}\eta_{3,5}/2$ to $\dot{\mathcal{N}}_{4,6}$.
Combining all these cases, we have
\begin{align}
   \dot{\mathcal{N}}_{i,j}
      = &- N_\mathrm{c} \eta_{i,j}         
          \underbrace{
          - \frac{\bar z N P_{i,j}}{2} \left( \sum_k \eta_{\langle i,k\rangle}
          - \sum_\ell \eta_{\langle \ell,i+2-\ell\rangle} \right)
          }_{= \frac{\bar z N P_{i,j}}{2} (\dot Q_i - Q_i \sum_{m,n} \eta_{m,n})} \notag \\
          &
          \underbrace{
          - \frac{\bar z N P_{i,j}}{2} \left(\sum_k \eta_{\langle j,k\rangle}
          - \sum_\ell \eta_{\langle \ell,j+2-\ell\rangle} \right)
          }_{= \frac{\bar z N P_{i,j}}{2} (\dot Q_j - Q_j \sum_{m,n} \eta_{m,n})} \ ,
\label{eq:eomsNij}
\end{align}
in which, for simplicity, $\langle i,j\rangle$, denotes ordering of the indices, $j>i$, with $\eta_{\langle i,j\rangle}= 0$ when $j<i$. This avoids double counting, obviating the $\delta$-functions and the $1/2$ factor in Eqs.~\eqref{eq:masterEqs}.
The first term in Eq.~(\ref{eq:eomsNij}) arises from case~(i) and the second and third arise from cases~(iia) and (iib), respectively. Only the indices $i$ and $j$ change between the second and third term.
In the second term, the reaction $\eta_{\langle j,k\rangle}$ depletes $\mathcal{N}_{i,j}$ when $i+k\to (i+k-2)$ and increases it when $(i+k-2)\to i+k$ (case~(iia)). 
The reaction $\eta_{\langle\ell,i+2-\ell\rangle}$ increases $\mathcal{N}_{i,j}$ when $\ell+(i+2-\ell)\to i$ and depletes it when $i\to \ell+(i+2-\ell)$ (case~(iib)). The third term affects $\mathcal{N}_{i,j}$ similarly.
Using Eq.~\eqref{eq:masterEqs}, the terms in the brackets of the second and third terms reduce to the expressions written below them.
Using then the relation~\cite{Wanjura2020}
\begin{align}
\sum_{\substack{m,n=3\\m\leq n\\m+n-2\leq C}} \hspace{-2.5ex}
\eta_{m,n}=-\frac{\dot{N}_\mathrm{c}}{N_\mathrm{c}}=-\frac{\dot{\bar z}}{\bar z-2} \ ,
\label{eq:zdot}
\end{align}
Eq.~(\ref{eq:eomsNij}) simplifies to
\begin{align}
   \dot{\mathcal{N}}_{i,j}
      = & - N_\mathrm{c} \eta_{i,j}
          + \frac{\bar z N P_{i,j}}{2} \left(\dot Q_i + Q_i\frac{\dot{\bar z}}{\bar z -2}\right) \notag \\
          & + \frac{\bar z N P_{i,j}}{2} \left(\dot Q_j + Q_j \frac{\dot{\bar z}}{\bar z -2}\right) \ .
          \label{eq:numberNeighboursDerivative}
\end{align}
At steady state, all the time derivatives vanish and Eq.~(\ref{eq:numberNeighboursDerivative}) yields $\eta_{i,j} = 0\ \forall\ i,j$,
which is the DBSS. We discuss this surprising result in the concluding section. 
We note that underlying relation (\ref{eq:neighbourProb}) is an assumption that $P_{i,j}(t)$ is uniform across the system. Consider then the conditional probability $Q_{i\mid j}$. Since $P_{i,j}$ is symmetric in $i$ and $j$, not only $Q_{i\mid j}=Q_{j\mid i}$, but this symmetry applies to all pairs of $j$- and $k$-cells around $i$. This leads to the conclusion that, overall, $Q_{i\mid j}=Q_i$, as has been indeed observed experimentally here and in~\cite{Sun2021}.

\section{Ramification for maximum-entropy CODs}

For systems with sufficiently high inter-particle friction and weak loading, quasi-static dynamics give rise to steady states CODs that maximise the entropy, $S\equiv-\sum_k Q_k \ln Q_k$~\cite{Sun2020}. The maximal entropy can be constrained either by the steady-state mean cell order $\bar{k}=\sum_{k=3}^{C}kQ_k$, $\bar{z}$, or by $Q_3$~\cite{Bl21pack}. Using $\bar{k}$, for example, the relevant Lagrangian is
\begin{align}
\mathcal{L} = & -\sum_{k=3}^{C}Q_k\ln{Q_k} - \mu\left(\sum_{k=3}^{C}Q_k - 1\right) \nonumber \\
&- \lambda\left(\sum_{k=3}^CkQ_k - \bar{k}\right) \ ,
\label{Entropy}
\end{align}
with $\mu$ and $\lambda$ the Lagrange multipliers corresponding to the normalisation and mean constraints, respectively. 
Maximising with respect to $\mu$ and imposing the normalisation condition, yields
\begin{equation}
Q_k = \frac{e^{3\lambda}\left(1-e^{-\lambda}\right)}{1-e^{-\lambda\left(C-2\right)} }e^{-\lambda k} \equiv Ae^{-\lambda k} \ .
\label{QkEnt}
\end{equation}
Imposing then the mean order condition, yields an algebraic equation for $e^{-\lambda}$,
\begin{equation}
\sum_{k=3}^{C} \left(k - \bar{k}\right) e^{-\lambda k} = 0 \ ,
\label{LambdaEnt}
\end{equation}
whose solution yields  $\lambda\left(\bar{k},C\right)$. The solution can also be expressed in terms of the mean coordination number $\bar{z}$ by using Euler's relation for planar graphs~\cite{Bletal15}
 \begin{equation}
\bar{k}\ = \frac{2\bar{z}}{\bar{z}-2} \ .
\label{Euler}
\end{equation}
Using the solution for $\lambda$ in Eq.~\eqref{QkEnt} gives the maximum-entropy COD for finite $C$, generalising the solution obtained in \cite{Sun2020}.

Substituting the maximum entropy COD, (\ref{QkEnt}), in the detailed balance condition $\eta_{i,j}=0$,
\begin{align}
\eta_{i,j} & = p_{i,j}Q_iQ_j - q_{i+j-2,i}Q_{i+j-2} \nonumber \\
& = Ae^{-\lambda\left(i+j-2\right)}\left(Ap_{i,j}e^{-2\lambda} - q_{i+j-2,1}\right) = 0 \ ,
\label{MEDB}
\end{align} 
we obtain
\begin{align}
\alpha_{i,j}\equiv\frac{p_{i,j}}{q_{i+j-2,i}} = \frac{e^{2\lambda} }{A} = \frac{1 - e^{-\lambda\left(C-2\right)}}{e^\lambda - 1} \ ,
\label{alpha}
\end{align}
i.e., the rates ratio in each process, $\alpha_{i,j}$, is independent of $i$ and $j$! This means that the COD, which is a key characteristic of the structure, is determined by this parameter alone. Thus, measurements of this parameter in experiments and simulations, as well as $Q_k$ and $\bar{k}$, can be used with eqs.~\eqref{QkEnt} and \eqref{LambdaEnt} to test our analysis.

\section{Experimental validation}

\begin{figure*}[htbp]
   \centering \includegraphics[width=\textwidth]{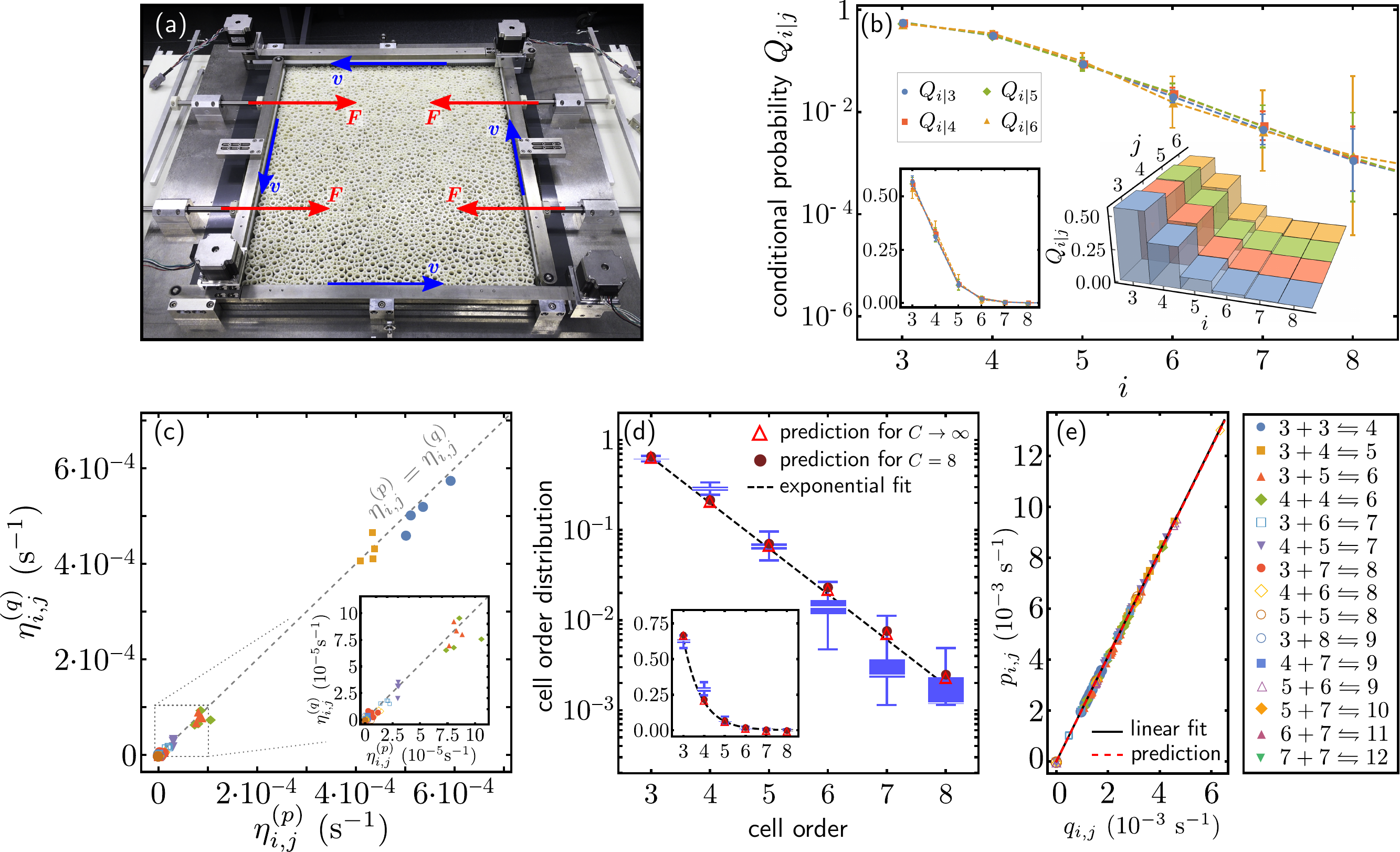}
   \caption{%
   \textbf{The rotational shear experiment.}
   (a)~The experimental setup -- an assembly of ABS particles is sheared by the motion of belts running at constant speed $v$ along the four boundaries (blue arrows). Weights attached via deflection pulleys maintain a constant force $F$ on two opposite sides (red arrows).
   (b)~The conditional COD -- the plots of $Q_{i\vert j}$ vs. $i$ collapse onto one curve for all $j$, showing that $Q_{i\vert j}$ is \emph{independent of $j$} and every cell sees the same neighbourhood independent of its order.
   (c)~The steady state reaction rates of merging, $\eta_{i,j}^{(p)}$, and breaking, $\eta_{i,j}^{(q)}$, balance out.
   (c) Individual data represent the average of $\eta_{i,j}^{(p)}$ and $\eta_{i,j}^{(q)}$ during one of the four measurement cycles for all the reactions up to, and including, cells order $C=12$).
   (d)~COD as a function of the cell order.
   We show the COD data in semi-log scale of $800\,\mathrm{s}$ of measurement as box plots and compare it to the COD we calculate from the experimental mean cell order $\bar{k}=3.48\pm0.02$ for $C=8$ (red disks) and $C\to\infty$ (red triangles). We also show an exponential fit as dashed lines. Theory prediction and experimental data are in good agreement. The fact that we slightly underestimate the fraction of $4$-cells in the system may be due to rattlers, or the relatively low friction coefficient which favours smaller cells over the less mechanically stable larger cells.
   The inset shows the COD on a linear scale.
   (e)~Rates of merging and breaking for different reactions.
   Each measurement is shown as individual data and the mean for each reaction is shown in the inset.
   The ratio is the same for all rates because all fall on the same line. The theory prediction for this ratio, eq.~\eqref{alpha}, is practically identically with a fit to the experimental data.}
   \label{fig:overview}
\end{figure*}

We tested the above results in rotational shear experiments, using the setup shown in Fig.~\ref{fig:overview}(a).
$2410$ Acrylonitrile butadiene styrene (ABS) rings of inter-particle friction coefficient~$\mu=0.46\pm0.05$ were assembled on a plate Four diameters were used to minimise crystallisation: $600$ of $7$mm, $600$ of $9$mm, $680$ of $11$mm, and $530$ of $14$mm.
Assemblies were sheared by confining peripheral belts, moving at a constant speed $v$ (blue arrows in Fig.~\ref{fig:overview}(a)). 
A constant force (red arrows in Fig.~\ref{fig:overview}(a)) was maintained on two opposite boundaries  by weights on deflection pulleys. The boundaries were free to move in and out to keep the confining pressure constant.
An overhead camera took one photograph per second and particles were tracked between frames, using an in-house Mathematica-based code. The code was used for each frame to find particle positions, identify contacts, obtain the contact network and cell structure, extract the COD, and, by comparing individual cell orders between consecutive frames, compute the rates $p_{i,j}$ and $q_{i,j}$. Further details are available in the supplemental material~\cite{SM}. We limited the analysis to a circular region of diameter $1084$ pixels, in which the particle velocity profile was proportional to the radius, yielding a constant shear rate~\cite{SM}.

The initial states were prepared by placing the particles arbitrarily within the rectangular area before attaching the weights. To break any contacts and remove incidental correlations, a hair dryer was used to blow air from different directions into the spaces between particles.
A confining normal stress of $65.4$N/m was then applied to the opposing boundaries and the assembly was sheared at $v=1$cm/s for $10$ minutes to generate the initial state.
From this state, we ran the belts at $v=0.2$m/s and a pressure of $p=32.7$N/m.
We found that the COD reaches a steady state, $\dot Q_k\approx0$, after approximately $60$ minutes. Therefore, we first applied shear for $60$min and then recorded the measurements for another $60$ minutes. Data was collected in the last $800$s of that period. We repeated this procedure four times for improved statistics.
The images were used to construct the COD and the conditional probabilities $Q_{i\vert j}$ at each time step. To obtain rates, we counted contact events over periods of $200$s, obtaining four data sets for each of the four runs, altogether $16$ data points for each reaction. From these data we determined the rates $p_{i,j}$, $q_{i,j}$, the values of the cell merging rates, $\eta^{(p)}_{i,j}\equiv p_{i,j}Q_iQ_j$, and the cell splitting rates, $\eta^{(q)}_{i,j}\equiv q_{i+j-2,1}Q_{i+j-2}$.

To check that $Q_{i\mid j}$ is independent of $j$, we first plot these as functions of $i$ (Fig.~\ref{fig:overview}~(b)) up to $i=8$ and $j=6$, where statistics are best. The curves for different values of $j$ indeed collapse onto one curve for all $j$. 
Next, we plot the values of $\eta_{i,j}^{(q)}$ against $\eta_{i,j}^{(p)}$ (Fig.~\ref{fig:overview}~(c)). We find that, to very good accuracy, $\eta_{i,j}^{(p)}=\eta_{i,j}^{(q)}$, namely, $\eta_{i,j}=0$ and these are indeed DBSS. 
The steady state COD is shown in Fig.~\ref{fig:overview}~(d), with error bars representing the $5\%$ to $95\%$ quantiles. 

To test if our measured COD maximises the entropy, we used the measured value of $\bar{k}=3.480\pm0.017$ in eqs.~(\ref{Entropy})-(\ref{alpha}) calculate the exponent of the maximum-entropy exponential COD, once for $C=8$ and once for $C\to\infty$.
For the former, we find $\lambda_8=1.116\pm0.026$ and for the latter, $\lambda=1.13\pm 0.03$.
Fitting the experimentally measured values of $Q_k$ to an exponential, we obtained $\lambda_\mathrm{exp} = 1.17\pm 0.10$. 
Thus, there is a good agreement between the measurements and the prediction for $C\to\infty$, except for a slight discrepancy at $Q_4$.
Measuring the number of merging and splitting events between $t$ and $t+\Delta t$, respectively, $n_{i,j}^{(p)}$ and $n_{i,j}^{(q)}$, we then calculated the rates, $p_{i,j}$ and $q_{i,j}$, using the fitted COD of Fig.~\ref{fig:overview}(e) in the normalised expressions
\begin{align}
   p_{i,j} & = \frac{n_{i,j}^{(p)}(t,\Delta t)}{Q_i Q_j N_c \Delta t},
   \quad\quad
   q_{i,j} = \frac{n_{i,j}^{(q)}(t,\Delta t)}{Q_{i+j-2} N_c \Delta t} \ .
   \label{eq:DefRates}
\end{align}
The data collapse nicely onto a straight line, indicating $\alpha_{i,j}=2.05\pm 0.08$ is a constant. Using the calculated value of $\lambda_8$ in eq.~\eqref{alpha}, yields $\alpha_{i,j}=2.062\pm 0.002$, in excellent agreement with the measurements. The negligible fit error originates from propagation of the assumed uncertainty in the rates. We discuss this test in more detail in the supplemental material~\cite{SM}. 

\section{Conclusion}

To conclude, we investigated theoretically and experimentally the emergence of detail-balanced steady states (DBSS) in planar granular dynamics. We have shown analytically that this is a direct consequence of the uniform spatial distribution of a neighbour probability, $P_{i,j}$. This leads to the conditional distribution around cells, $Q_{i\mid j}$, being independent of $j$, which has been observed in our experiments and in~\cite{Sun2021}. 
Our analysis and results suggest that DBSS may be more common than previously thought, as idea that would be interesting to check under different conditions.

We also showed that, when the steady-state COD also maximises the entropy~\cite{Sun2020}, then the merging-to-splitting ratio in all the processes, $p_{i,j}/q_{i+j-2,i}$, is the same.

Two aspects make these results significant. One is that, in view of the enormous impact that the detailed balance principle has had in equilibrium physics models~\cite{Ma1867,Bo1872,Ein17,Di24,On31}, it can similarly open the door to better models in granular dynamics. E.g., by including considerations of entropy and stability and, in particular, in dynamics that maximise the COD's entropy.  
More generally, this understanding of the origin of DBSS in our systems can guide searches for this phenomenon in other non-equilibrium processes.

\section{Acknowledgements}
We thank Karolina Zeh and Martin M\"uller for assisting with designing and building the experimental setup.

\appendix
\clearpage
\twocolumngrid
\begin{center}
\widetext
{\large\bf
Detailed balance in non-equilibrium dynamics of granular matter: \\
derivation and implications \\
---Supplementary Material---} \\[\baselineskip]
\end{center}
\twocolumngrid

\setcounter{equation}{0}
\setcounter{figure}{0}
\setcounter{table}{0}
\setcounter{page}{1}

\renewcommand{\theequation}{S\arabic{equation}} % prepend 'S' to equation numbers
\renewcommand{\thefigure}{S\arabic{figure}}

\begin{figure*}[htbp]
   \centering
   \includegraphics[width=.99\textwidth]{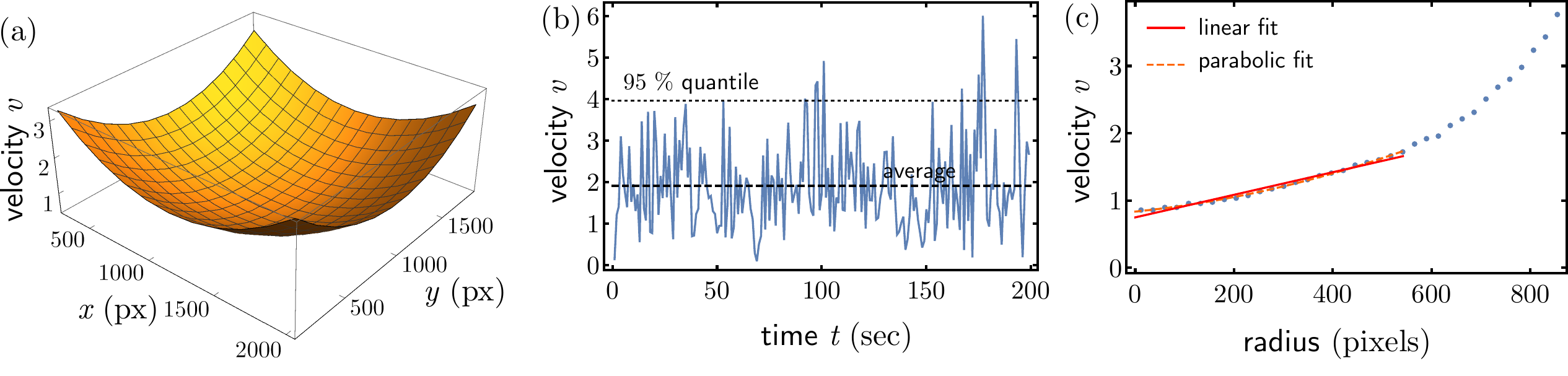}
   \caption{\textbf{Particle AVCs and the region of constant shear rate.}
   (a)~We obtain the system's centre through a parabolic fit to the AVC data.
   (b)~Representative example of a particle AVC. Characteristically, particles move in jolts rather than at constant speed. The $95\,\%$ quantile represents the maxima of these peaks and it is these values that we relate to the shear rate, rather than the mean.
   (c)~$95\,\%$ quantile of the velocity as a function of the distance to the system centre (radius). Up to a radius $\leq546\,\mathrm{pixels}$, there is a linear regime, in which the parabolic and linear fits hardly differ.
   }
   \label{fig:velocityRadius}
\end{figure*}

\nin\textbf{Subsystem with constant shear rate:}
Since the experimental system is sheared on all sides, the particles follow approximately circular trajectories and their velocities are rotationally symmetric with respect to the centre. We only analyse the part of the system with approximately constant shear rate / azimuthal velocity component (AVC). To identify this regime, we analysed the particle AVCs at different radii from the centre. 
First, we obtained the centre by fitting the function
\begin{align*}
   v(x,y) = & v_0 + a ((x - x_0)^2 + (y - y_0)^2) \\
            & + b\sqrt{(x - x_0)^2 + (y - y_0)^2} + c
\end{align*}
to the 2 dimensional velocity data $v(x,y)$. To this end, we used the data from all time frames. This function is rotationally symmetric with respect to $(x_0,y_0)$, which is the centre of the rotation, see Fig.~\ref{fig:velocityRadius}(a).

Next, we analysed the AVC as a function of time. The AVC of a particle does not stay constant, but rather, particles are accelerated and decelerated in response to changes in the local forces and the structure, see Fig.~\ref{fig:velocityRadius}(b). It is plausible that this velocity's maximum is correlated with the shear rate. Since the maximum is relatively noisy, we used the $95\,\%$ quantile of the AVC for our analysis and we plot it as a function of the radius in Fig.~\ref{fig:velocityRadius}(c). The plot substantiates that there is a regime $0\leq r\leq r_0=542$ pixels over which $v(r)$ is linear. In this region $\partial v(r)/\partial r\equiv\mathrm{constant}$ and it is therefore this region that we used for our analysis.\\

\begin{figure*}[htbp]
   \centering
   \includegraphics[width=\textwidth]{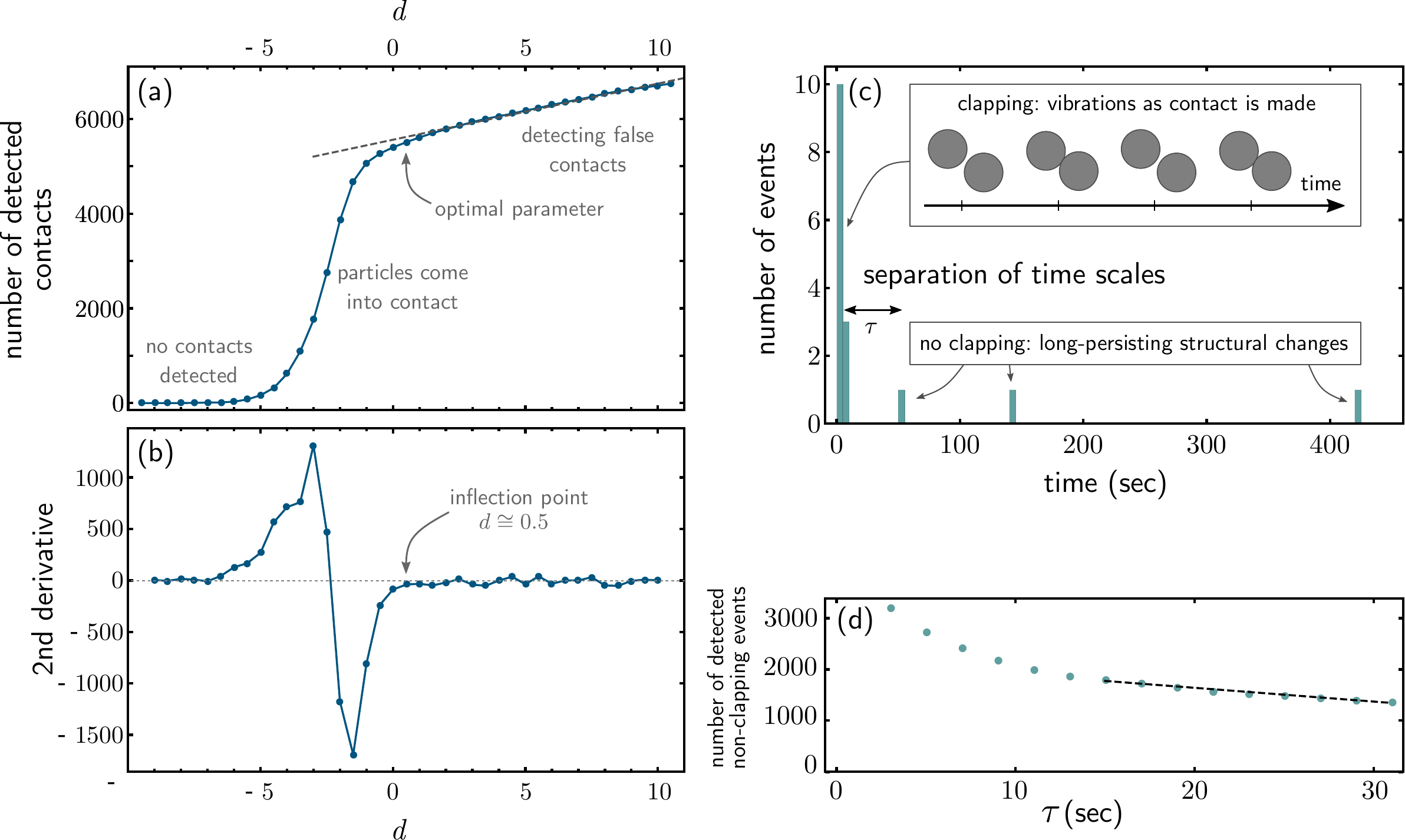}
   \caption{\textbf{Contact detection and detecting structure-changing contact events.}
   (a)~The number of detected contacts as a function of the detection parameter $\mathbf{d}$.
   The sharp increase in the number of particles, when they start coming into contact, is followed by a slow, approximately linear, increase (grey dashed line). The slow increase is the result of falsely detected contacts. The optimal choice of $d$ is immediately to the right of the sharp increase, just before the linear increase and when the second derivative shown in~(b) vanishes. This value is $d=0.5$.
   %\textbf{Clapping vs. non-clapping events.}
   (c)~Residual kinetic energy can result in vibrations of particles, causing repeated making and breaking of a contact between them. Clapping takes place typically on a time scale $\leq\tau$ that is sorter than non-clapping events. 
   (d)~The number of detected events drops rapidly with and then settles into a linear decrease beyond  $\tau=15$sec. This is when the number of non-clapping contacts start to be filtered out and therefore is the optimal value for filtering out only clapping events.
   }
   \label{fig:contactDetection}
\end{figure*}

\nin\textbf{Contact detection:}
Two particles, $j$ and $k$, were considered in contact when their centroids, $(x_{j,k},y_{j,k})$, and radii, $r_{j,k}$, obeyed
\begin{align}
   (x_j-x_k)^2 + (y_j - y_k)^2 \leq (r_j + r_k + d)^2
\end{align}
with $d$ an additive constant that effectively enlarges the particle radius to make them overlap.
The optimal value of $d$ was determined by running the image analysis for different values of $d$ and counting the number of detected contacts in the system. Increasing $d$, particles start coming into contact at some value $d_c$ and the number of detected contacts increases sharply. Thereafter, the number of contacts increases more slowly (and seemingly linearly), as more and more false contacts are detected. This is shown in Fig.~\ref{fig:contactDetection}(a). We defined the optimum value of $d$ slightly to the right of the initial sharp increase in the number of contacts to eliminate the error originating from the detection of false contacts. Practically, this means that we chose the smallest value of $d$ for which the second derivative of the number of detected contacts vanishes, see Fig.~\ref{fig:contactDetection}(b). This value is $d=0.5$.

\nin\textbf{Detecting events and clapping:}
As reported in simulations~\cite{Wanjura2020}, pairs of particles can `clap', namely, make and break contact repeatedly. We also observed clapping in these experiments. This phenomenon does not change the balance of reactions and therefore neither affects the observation of the detailed balance in steady states, $\eta_{i,j}=0$ $\forall\ i,j$. Clapping takes place in the experiment on relatively well-defined time scales.
On making contact, momentum can drive the two particles slightly apart until the compressive force field pushes them into contact again. This can be repeated until this kinetics is damped. This separation of scales allowed us to filter out such events by removing any events involving the same two grains that happen within a certain time duration $\tau$. In our analysis, when a pair of particles clap during any time shorter than $\tau$, this is counted as one event if the number of claps is odd, in which case the final structure has changed. Otherwise the contact state of the two particles is deemed unchanged. 
An illustration of this process is shown in Fig.~\ref{fig:contactDetection}(c). Initially, particles make and break contact rapidly and eventually settle. Later events are well separated in time from the initial clapping.
To determine $\tau$, we plot the number of detected events as we vary $\tau$ (Fig.~\ref{fig:contactDetection}(d)). This function decreases faster than linearly when clapping is involved and transitions into a linear decrease when non-clapping events starts to be filtered out. The transition is at $\tau=15$sec, which is therefore the optimal value for filtering out only the clapping events.  \\

\begin{figure}[htbp]
   \centering
   \includegraphics[width=.49\textwidth]{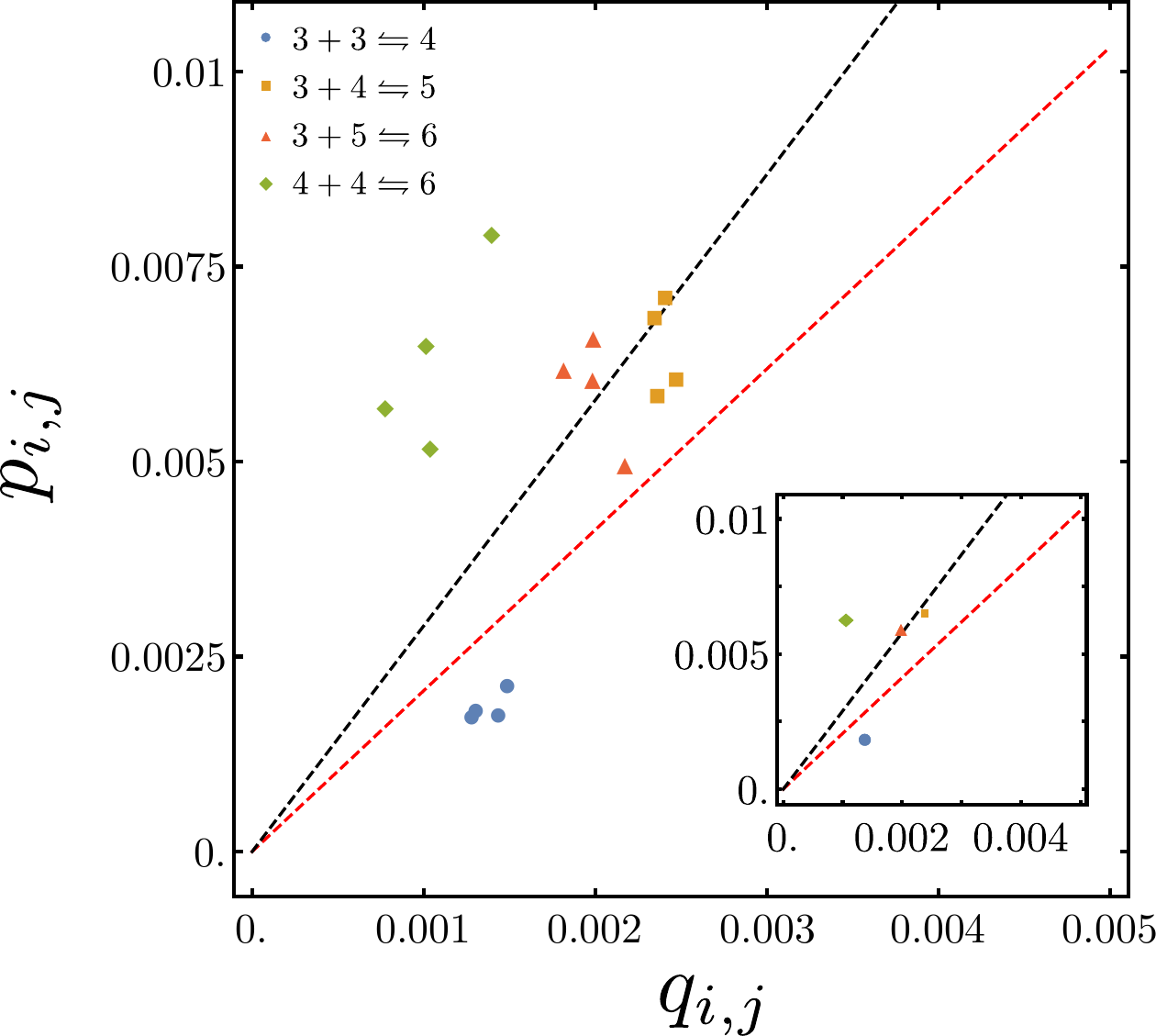}
   \caption{\textbf{Rates ratios calculated from COD data.}
   Rather than using the fitted COD shown in Fig.~\ref{fig:overview}(h), we divide by the mean COD during the respective time interval during which we count the contact events. This leads to a broader distribution of the data, and a discrepancy between a linear fit to the data (black dashed line) and the theory (red dashed line) due to rattlers and boundary effects.}
   \label{fig:honestRatesRatios}
\end{figure}

\nin\textbf{Rates ratios:}
We calculate the rates from the number of merging $p_{i,j}$ and breaking $q_{i+j-2,i}$ events $n_{i,j}^{(p/q)}$, see eq.~\eqref{eq:DefRates}.
The values of the fractions, $Q_k$ were obtained from the exponential fit, shown in Fig.~2~(d) in the main text rather than directly from the cumulative experimental data during $\Delta t$, which includes unavoidably large fluctuations. These large fluctuations propagate to the rates and are amplified due to the division by $Q_k$. For completeness, we provide the plot with reactions up to $C=6$ in Fig.~\ref{fig:honestRatesRatios}.

\begin{thebibliography}{99}
\bibitem{footnoteCycles} The only possible reactions for $k<6$ which cannot give rise to cycles according to the master equations~\eqref{eq:masterEqs} are $3+3\leftrightharpoons{}4$ and $3+3\leftrightharpoons{}5$. For higher maximal cell orders, cycles are possible. For instance, for $k<7$, we have $3+3\leftrightharpoons{}4$, $3+4\leftrightharpoons{}5$, $3+5\leftrightharpoons{}6$ and $4+4\leftrightharpoons{}6$, which, in principle, would admit the cycle $3+5\to 6\to 4+4 \to 6$.
\end{thebibliography}
\end{document}